\documentclass[12pt]{article}

\usepackage[utf8x]{inputenc}
\usepackage[english]{babel}

\usepackage{amsmath}
\usepackage{epsf}
\usepackage{epsfig}
\usepackage{amssymb}
\usepackage{graphicx}
\usepackage{multirow} 
\usepackage{multicol}

\def\k{{\bf k}}

\begin{document}

\title{ \bf  Probing millicharged particles with NA64 experiment at CERN}

\author{  S.N.~Gninenko$^{1}$\footnote{{\bf e-mail}: sergei.gninenko@cern.ch},
 D.V.~Kirpichnikov$^{1}$\footnote{{\bf e-mail}: kirpich@ms2.inr.ac.ru},
    and N.V.~Krasnikov$^{1,2}$\footnote{{\bf e-mail}: nikolai.krasnikov@cern.ch}
\\
\\
 $^{1}$ Institute for Nuclear Research of the Russian Academy of Sciences, \\117312 Moscow, Russia \\
 $^{2}$ Joint Institute for Nuclear Research, 141980 Dubna, Russia}


\date{\today}

\maketitle

\begin{abstract}
In this note  we estimate the sensitivity of the NA64 experiment to millicharged particles ($\chi$). That experimental facility is dedicated
to the searching for dark sector particles in missing energy events  at the CERN SPS. We consider missing momentum signatures in the $\simeq$ 100 GeV
  electron and muon beams  
  and show  that the later one  allows to obtain more stringent 
bounds on the millicharge $Q_{\chi}$, which  for the $\chi$  masses   
$100$~MeV $\leq m_{\chi} \leq 500$~MeV  at the level $ Q_{\chi}/e\lesssim O(10^{-3}) - O(10^{-2})$.

\end{abstract}


\section{Introduction}
The millicharged particles ($\chi$), i.e. particles  with an electric charge $Q_{\chi}=\epsilon e$ much  smaller ($\epsilon \ll 1$) than the  elementary charge $e$, 
have been considered long ago. Those particles were discussed in connection with  the mechanism of the electric charge quantization and  a possible nonconservation of the electric charge \cite{ignatiev}. 
In the context of grand unification models this  mechanism may be linked  to  
magnetic monopole and    electric charge quantization~\cite{Dirac:1931kp}.  
 However, the magnetic monopoles 
have not been observed yet, and the underlying mechanism for charge 
quantization remains non confirmed, thus making  searches for millicharged particles  of a great interest. 
\par The  $\chi$s have been also proposed in various extensions  of the Standard Model (SM). 
In particular, in the hidden (dark) sector models with a new $U'(1)$ gauge group  \cite{Holdom:1985ag}, see also \cite{Okun:1982xi}. 
In this scenario   the kinetic mixing between 
the  $U'(1)$  and the SM fields   is described by a term $ \frac{\epsilon}{2} F_{\mu \nu}' F^{\mu \nu} $. 
Where $ F_{\mu \nu}'  = \partial_{\mu}A'_{\nu} - \partial_{\nu}A'_{\nu}$ 
and  $ F_{\mu \nu}  = \partial_{\mu}A_{\nu} - \partial_{\nu}A_{\nu} $ with  
$A'_{\mu}$ and $A_{\mu}$ being a dark photon and ordinary photon  
respectively.
 After a redefinition of the hidden vector field $A_\mu' \rightarrow A_\mu' +  \epsilon A_\mu$
  one can find that electromagnetic field $A_{\mu}$ interacts with hidden fermions of dark sector, namely 
  $\mathcal{L}_{int}  =   \epsilon A_\mu J^\mu_{D}$. Here  
$J_D^{\mu} = g_D \bar{\chi} \gamma^\mu \chi $ is the $U'(1)$  current of the  fermions from dark sector. This means that
the dark sector  particles $\chi$  interact with the photon via the effective coupling 
$Q_\chi =  \epsilon g_D$.  
In this scenario the   dark photon remains  massless and interacts only with dark sector particles.
In the rest of this work,  we will consider the following  Lagrangian for  the  $\chi$s   interacting with the 
electromagnetic field $A_{\mu}$, assuming that they are  spin $1/2$ fermions:  
\begin{equation}
\mathcal{L}  \supset  i \bar{\chi} \gamma^\mu \partial_\mu \chi  
- m_\chi \bar{\chi} \chi +
  Q_\chi A_\mu \bar{\chi} \gamma^\mu \chi \,.
\label{AChiChi}
\end{equation}
where $m_\chi$ is a Dirac mass of the hidden particles. 

As it follows from \eqref{AChiChi} in  the leading order the $\chi$ 
production rate is proportional to  $Q_\chi^2$ and the $\chi$s  can be effectively produced  in  any electromagnetic  reactions if  kinematically allowed \cite{ignatiev1}.
The numerous constraints on $Q_\chi$ obtained from the dedicated  beam-dump  \cite{golow, Prinz:1998ua},  positronium  \cite{bader} and reactor  \cite{gkr, singh} experiments,
 see  Ref. \cite{david3} for a review. The  expected limits from the   
  $e^+e^-$ colliders and  the LHC \cite{Izaguirre:2015eya, Liu:2018jdi} have been reported recently. 
 Stringent constraints on $\chi$ can also be obtained from cosmological and astrophysical considerations, see e.g. \cite{gold, david1, david2, david3, dub, dolg, zurab1, zurab2}.  
\par The millicharged particles with $Q_\chi \ll e$   typically escape the detection in an  experiment\footnote{Here $e$ is the electron electric charge, 
$\frac{e^2}{4\pi} = \frac{1}{137}$.}, because their ionisation energy loss is $\sim Q_\chi^2$ and thus is very small.  
Therefore,   to observe them directly a large number of particles on target is required, see e.g. ~\cite{Prinz:1998ua,Magill:2018tbb}. 
However,  possible indirect observation of  $\chi$s  at the fixed-target facilities  can utilise  another  more effective approach -
 the search for the $\chi$s in missing energy/momentum events \cite{sng, Gninenko:2016kpg, Berlin:2018bsc, Akesson:2018vlm}. 
 
Let us consider the  NA64 experiment at CERN ~\cite{Gninenko:2016kpg,Banerjee:2016tad,Banerjee:2017hhz,Gninenko:2017yus}, which was 
designed to search for the  light dark matter particles in the  reaction of dark photon production $eZ \rightarrow eZ A'$ followed by  the 
invisible decay of dark photon into hidden states, $A' \rightarrow invisible$. However,  the missing energy  signature 
for the search of dark photons  can also be implemented to search for the millicharged particles produced in the similar 
reaction  $eZ \rightarrow eZ \chi \bar{\chi}$. 
At present NA64 experiment uses the electron beam with the energy $E_0 \approx 100$~GeV, but 
there are also plans to use  the high intensity muon M2  beam line  at the Super Proton Synchrotron (SPS)  at 
CERN \cite{addendum}. 
Moreover, the  missing momentum experiments with muon beams  at CERN~\cite{Gninenko:2014pea,Gninenko:2018tlp,Chen:2018vkr}  and  
FermiLab	~\cite{Kahn:2018cqs,Chen:2017awl}
have been proposed recently in order to 
probe $(g-2)_\mu$ anomaly~\cite{Bennett:2004pv} in the framework
of light dark matter sector~\cite{Berlin:2018bsc,Akesson:2018vlm}.
  
It should be noted that the  LHC experiments 
are insensitive to probe sub-GeV dark sector scenario~\cite{CMS:2012xi} with small coupling constants. 
In particular, the millicharge parameter space  
  in the ranges $0.1$ GeV $\lesssim m_\chi \lesssim 1$ GeV and 
  $10^{-4} \lesssim Q_\chi /e \lesssim 10^{-3}$ has not 
 been constrained yet by existing  experiments. 
The  scenarios of sub-GeV hidden particles can be probed  at 
SHIP~\cite{Anelli:2015pba} proton 
beam dump  facility as well as at MiniBoone~\cite{Aguilar-Arevalo:2018gpe}, DUNE~\cite{Acciarri:2015uup} and LSND~\cite{Athanassopoulos:1996ds} neutrino 
detectors.  In these experiments
the dominant millicharge production signatures are  exotic  decays
\begin{equation}
   \pi^0/ \eta \rightarrow \gamma \chi \bar{\chi}, \qquad J/\psi, \Upsilon \rightarrow \chi \bar{\chi}. 
 \end{equation} 
 The produced millicharged particles
 elastically scatter on an atomic electrons in the dump, 
 $\chi e \rightarrow \chi e$. So the 
 detection of millicharged particles  is based on the measurement
 of low  energy electron recoils. The millicharge 
  yield from  hadrons~\cite{Magill:2018tbb} is suppressed 
by both the  production term $\sim Q_\chi^2$ and the interaction 
factor~$\sim Q_\chi^2$, such that $N_{\chi \chi} \sim Q_\chi^4$. 
On the other hand, the number of produced millicharged particles at NA64 
is proportional to  $ Q_\chi^2$ for both electron and muon beams.
For muon beam  significant gains in millicharge sensitivity compared to electron beam may be 
achieved by optimizing the active target design of NA64. In particular,  $10^{13}$ muons on 
target  are expected to accumulate at NA64
 during the couple of months running. 

In this note we estimate the discovery potential of millicharged particles  
at NA64 experiment for both electron and muon beams. We find that 
 muon beam setup of NA64  provides  more 
stringent bounds on electric charge $Q_{\chi}$
in comparison with electron beam.
The main reason is that  $100$-GeV electron beam degrades significantly even in 
 the relatively thin lead target of 40$X_0$  ($\approx$ 20 cm) used at NA64
\footnote{Here $X_0$ is electron radiation length}. 
 Therefore,
 the  electron missing momentum yield is suppressed by the 
electron beam attenuation factor $X_0$ and the number of produced millicharged particles 
is proportional to 
 $N_{\chi \chi} \sim X_0$. On the other hand
muon radiation length is $X^{\mu}_0 \sim (m_{\mu}/m_e)^2 X_0 \gg X_0$, thus the
relativistic $100$ GeV muons pass through the  dump with $L \ll X^{\mu}_0$
without significant loss of muon energy. 
This implies that the  millicharge production  signal in the muon beam experiment 
is proportional to the length of the target, $N_{\chi \chi} \sim 40 X_0$.\footnote{Here as an estimate  we use the length of target for muon experiment 
$L \sim 40 X_0  \approx 20~cm$.} Furthermore, one can
 improve the millicharge sensitivity for the muon beam by increasing  the 
 effective  interacting 
 length of the  active lead target. For instance, by increasing the length
 of the target by 4 factor of magnitude,  one can   
  extend  $Q_{\chi}$  bound by factor 2. 
 This provides 
 an excellent opportunity for NA64 with muon beam  to probe wider range of 
 millicharge  parameter space. We also derive NA64 bound on millicharges from 
recent NA64 experimental bound on $\epsilon$ parameter~\cite{Banerjee:2017hhz}
 for dark photon model. 

The organization of paper is as follows. In Section~\ref{BASIC} we collect basic formulae which are
relevant  for an estimation of the millicharged particles production 
rates. In Section~\ref{Elimits} we present bound on $(Q_\chi, m_\chi)$ from 
NA64e experiment. In Section~\ref{MillBounds} we discuss 
expected limits from combined analysis of NA64$e$ and NA64$\mu$. Last section summarises 
the main results.  

\section{The cross sections
\label{BASIC}}

\begin{figure}[tbh!]
\begin{center}
\includegraphics[width=0.9\textwidth]{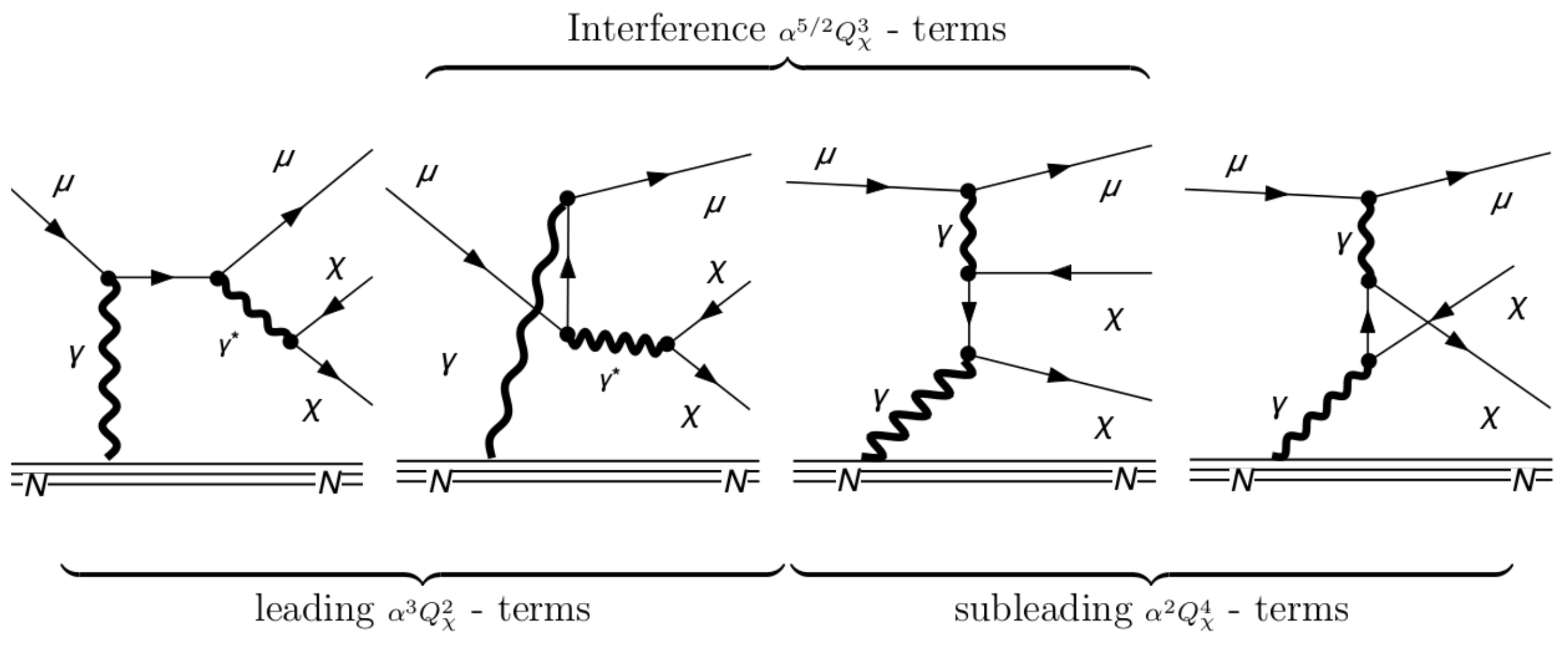}
\end{center}
\caption{Feynman diagrams of millicharge pair production} 
\label{FeynDiag}
\end{figure}

In this section we present basic formulae for  the cross-section of the high-energy lepton scattering on heavy nuclei 
accompanied by the emission of a bremsstrahlung $\chi \bar{\chi}$ pair
\begin{equation}
l N \rightarrow l  N \gamma^* 
\rightarrow l N \chi \bar{\chi}
\label{eNToeNchichi}
\end{equation}
with $\chi$ being a millicharged Dirac fermion and $l = e, \mu$.
 The relevant tree level diagrams are shown in  Fig.~\ref{FeynDiag} for
 muon case.  
The dominant millicharge emission cross-section  
(\ref{eNToeNchichi}) is proportional to $\mathcal{O}(\alpha^3 Q_\chi^2)$. 
We neglect $\mathcal{O}(\alpha^2 Q_\chi^4)$ 
trident millicharge production cross section and relevant 
$\mathcal{O}(\alpha^{5/2}Q_\chi^3)$ interference terms
 in our calculations. Therefore,
the differential  
cross-section $\sigma( l N \rightarrow l N \chi \bar{\chi})$ can be represented in the form~\cite{Beranek:2013yqa}
\begin{equation}
d \sigma(l N \rightarrow l N \chi \bar{\chi})=
d \mbox{Lips}_{2 \rightarrow 3} 
\overline{|\mathcal{M}_{2 \rightarrow 3}|^2}_{\alpha \beta} 
\frac{d k^2_{\gamma^*}}{(2 \pi)} \times  \chi_{\alpha \beta}
\end{equation}
where $d{\mbox{Lips}}_{2\rightarrow 3}$ is Lorentz invariant phase space
for a process $l N \rightarrow l N \gamma^*$ with off-shell photon 
in the final state, 
\begin{equation}
\overline{|\mathcal{M}_{2 \rightarrow 3}|^2}_{\alpha \beta}  = 
\overline{\sum_{spin}} \mathcal{M}^\mu \mathcal{M}^{\nu \dagger} 
g_{\mu \alpha} g_{ \nu \beta} \cdot
\frac{1}{ k^4_{\gamma^*} }, 
\label{M2to3}
\end{equation}
where an averaging over initial lepton spin and summation over outgoing lepton state is performed.  
The millicharged tensor $\chi_{\alpha \beta}$
 has the following form
\begin{equation}
\chi_{\alpha \beta} = \int  \frac{d^3 \k_1}{(2\pi)^3 2 E_1} \frac{ d^3 \k_2}{(2\pi)^3 2 E_2}  (2\pi)^4 \delta^{(4)}(k - k_1 - k_2)
\sum j_\alpha j^*_\beta,
\label{mQCTens1}
\end{equation}
with $j_\alpha = Q_\chi \overline{\chi} \gamma_\alpha \chi$ being a 
millicharged current, $k\equiv k_{\gamma^*}$ is  a total four-momentum
of the millicharged pair, $k_{\gamma^*}=k_1+k_2$.  
This implies that the  millicharge production cross-section  $d \sigma(l N \rightarrow l N \chi \bar{\chi})$
can  be represented in the form form
\begin{equation}
d \sigma(l N \rightarrow l N \chi \bar{\chi}) =
d \sigma(l N \rightarrow l N \gamma^*) \times
\frac{Q^2_\chi}{12 \pi^2} \frac{d k_{\gamma^*}^2}{ k_{\gamma^*}^2}
\sqrt{1-\frac{4m_{\chi}^2}{k_{\gamma^*}^2}} \left(1+  \frac{ 2m_{\chi^2}}{k_{\gamma^*}^2}\right).
\label{diffChiProdCS1}
\end{equation}

The first factor in (\ref{diffChiProdCS1}) can be calculated in the 
equivalent photon 
approach~\cite{Liu:2017htz,Bjorken:2009mm}, the corresponding differential cross-section is  
\begin{equation}
 \frac{d}{dx} \sigma_{2\rightarrow 3} \approx \frac{2}{3}
 \frac{\alpha^3 \zeta |\k_{\gamma^*}| }{x \tilde{u}^2 E_0}
 \left[ m_l^2 x (-2+2x+x^2)-2 (3 - 3x+ x^2)\tilde{u} \right], 
 \quad x=E_{\gamma^*}/E_0
 \label{dsdxIWW1}
\end{equation}
with $\zeta$ being the photon flux from nucleus
\begin{equation}
\zeta = \int\limits_{ k_{\gamma^*}^4/(4E_0^2)}^{k_{\gamma^*}^2+m_l^2}  
\frac{dt}{t^2} \left[t-\frac{k_{\gamma^*}^2}{4E_0^2}\right]
\cdot Z^2 \left(\frac{a^2 t}{a^2 t+1}\right)^2 \frac{1}{(1+t/d)^2},
\label{photonFluxFF}
\end{equation}
where $a=111 Z^{-1/3}/m_e$ parametrizes the electron screening effect and $d = 0.164 A^{-2/3}$ GeV$^2$ stands to account the finite nuclear size. 
Such form-factor parametrization~(\ref{photonFluxFF}) accounts for elastic 
scattering effects only. The inelastic form-factor is proportional to $\sim Z$ and thus can be neglected in high-$Z$ target experiment.     
The quantities $\tilde{u}$ and $|\k_{\gamma^*}|$ in 
(\ref{dsdxIWW1}) are defined by 
$\tilde{u}=-k_{\gamma^*}^2 (1-x)/x-m_l^2 x$ and 
$|\k_{\gamma^*}|=(x^2 E_0^2-k_{\gamma^*}^2)^{1/2}$ respectively.
Therefore, one can estimate the $\chi \bar{\chi}$-production rate
by integrating (\ref{diffChiProdCS1}) over $\gamma^*$ invariant mass 
\begin{equation}
\sigma_{lN \rightarrow l N \chi \bar{\chi}}\approx 
\int\limits_{0.5}^1 dx\,
\int\limits^{y_{max}}_1 \frac{d y}{y} \, \sqrt{1-\frac{1}{y}}\left(1+\frac{1}{2y}\right)\times \frac{Q_\chi^2}{12 \pi^2}
\times \frac{d \sigma_{2 \rightarrow 3}}{dx}, 
\label{eNtoeNcc2}
\end{equation}
where we denote $y=k_{\gamma^*}^2/(4 m_\chi^2)$ and $y_{max}=x^2E_0^2/(4m_\chi^2)$. Here the lower limit in the integration 
over  $x$ corresponds to  the following  missing energy cut, 
$E_{miss}/E_0\equiv E_{\gamma^*}/E_0 > 1/2$. It is instructive 
to obtain approximate expression for~(\ref{eNtoeNcc2}). Indeed,
 the integral over $x$ in~(\ref{eNtoeNcc2})
can be estimated in a way analogous to that performed in~\cite{Bjorken:2009mm}. Namely, 
for $k_{\gamma^*} > m_l$ the integral saturates at $x\approx 1$, therefore
 one has 
\begin{equation}
\int\limits_{0.5}^1 dx\,  \frac{d \sigma_{2 \rightarrow 3}}{dx}
\approx \frac{4}{3} \frac{\alpha^3 \zeta }{(k_{\gamma^*})^2} 
\left[ \ln \frac{1}{2} \left(\frac{k_{\gamma^*}}{ m_l}\right)^2 + 
\mathcal{O}(1) \right],
\end{equation}
where $k_{\gamma^*} / m_l$  under the logarithm
 regulates a soft lepton singularity. The integration over $y$
 can be performed in the leading logarithmic order. So that  
 for $m_\chi \gtrsim m_l$ the cross-section (\ref{eNtoeNcc2})
 can be approximated with a logarithmic accuracy as
\begin{equation}
 \sigma(l N \rightarrow l N \chi \bar{\chi}) \approx 
 \frac{4}{3} \frac{\alpha^3 \zeta }{(2 m_\chi)^2} 
\left[ \ln \frac{1}{2} \left(\frac{2 m_{\chi}}{ m_l}\right)^2 + 
\mathcal{O}(1) \right] 
\times  \kappa \frac{Q_\chi^2}{12 \pi^2}, 
  \label{TotCSFactorized}
\end{equation}
 where $\kappa$ is a function which depends weakly on
 $m_\chi/E_0$.
Namely, for
 $m_\chi/E_0  \ll 1$, one has $y_{max}~\gg~1$, so that the  integral for $\kappa$ can be 
calculated straightforwardly
\begin{equation}
\kappa \approx \int\limits^\infty_1 \frac{dy}{y^2} \sqrt{1-\frac{1}{y}}\left(1+\frac{1}{2y}\right) =\frac{4}{5}.
\end{equation}
In addition, we note that (\ref{TotCSFactorized})
 generally resembles, up to the numerical factor $\sim Q^2_\chi$ and additive correction to the logarithm,
 the total cross-section for the  dark photon (DP) production~\cite{Bjorken:2009mm},
in which the dark photon  mass is redefined as $m_{A'} \rightarrow 2 m_\chi$. 
  This observation allows one to estimate analytically the expected
 constraints for the parameter space of the  millicharged particles  
 directly  for muon and electron beam at NA64.  
\section{Limits from NA64e
\label{Elimits}}  

One can link the bound on $\epsilon$ parameter for 
the model with dark photon and bound on the millicharge. Indeed, the interaction of 
dark photon with the SM particles has the form~\cite{Bjorken:2009mm}
\begin{equation}
\mathcal{L}_{dark} = \epsilon e J^{\mu}_{SM}A'_{\mu}
\end{equation}
where $J^{\mu}_{SM} = \frac{2}{3}\bar{u}\gamma^{\mu}u -\frac{1}{3}\bar{d}\gamma^{\mu}d 
- \bar{e}\gamma^{\mu}e + ... $ is the SM electromagnetic current, $\frac{e^2}{4\pi} = \frac{1}{137}$, 
$A'_{\mu}$ is the dark photon field 
and $\epsilon$ is unknown parameter.  The goal of the experiments is to derive the bound 
on $\epsilon$. On the other hand, the bound on $\epsilon$ depends on the dark photon mass $m_{A'}$. The 
cross-section of dark photon electro-production $\sigma(eN \rightarrow eNA')$  is 
proportional to the cross section of virtual photon electro-production, 
namely~\cite{Bjorken:2009mm} 
\begin{equation}
 \sigma(eN \rightarrow e N A') = \epsilon^2 \sigma(eN \rightarrow eN \gamma^*) 
\approx 
 \frac{4}{3} \frac{\alpha^3 \epsilon^2 \zeta }{m_{A'}^2} 
\left[ \ln \frac{1}{2} \left(\frac{m_{A'}}{ m_e}\right)^2 + 
\mathcal{O}(1) \right],   
\label{DarkPhotonCS}
\end{equation}
where $m^2_{A'} = k^2_{\gamma^*}$ is the four momentum
squared of virtual photon.

In a recent NA64e analysis~\cite{Banerjee:2017hhz} a  stringent 
experimental constraints on the dark photon 
coupling, $10^{-5} <\epsilon < 10^{-2}$, for the mass range 
$m_{A'}\lesssim 1$~GeV were derived. These bounds were 
obtained by using {\tt GEANT4} Monte Carlo simulation for
the flux and spectra of the $A'$s produced in the target by primary 
electrons.  That numerical simulation
 takes into account the development of the signal 
 electromagnetic (EM) shower in the reaction $eN\to eN A'\to eN \chi\chi$ in the target. 
We emphasize that the dark photons and millicharged pairs are 
 essentially produced within the first
radiation length of the target material for electron beam. 
Therefore, the beam attenuation and EM shower development should be taken into account in a proper 
numerical simulations of $\chi$-pair production. 
The latter is beyond of 
the scope of our paper.  We leave this task for future analysis.  
Instead, in order to obtain $95 \%$ C.L. bound on $Q^{(e)}_\chi$   
 we link it with an upper bound $\epsilon^2$ from Ref.~\cite{Banerjee:2017hhz}  as
\begin{equation}
\epsilon^2 \approx \kappa \cdot\frac{(Q_\chi^{(e)})^2}{12 \pi^2} ,
\label{mQBoundFromDP}
\end{equation}
that also implies
the replacement $m_{A'} \to 2m_\chi $ in the relevant exclusion plot.
Indeed, one can easily obtain~(\ref{mQBoundFromDP}) from Eqs.~(\ref{TotCSFactorized})
and (\ref{DarkPhotonCS}). 
We expect that  $\epsilon^2$ limit
 in  (\ref{mQBoundFromDP}) takes into account the 
processes  
associated with EM shower development   
in the  thick target for the same kinematical cuts of energy deposition.  
In particular,  in the present analysis
 the missing  energy cut~\cite{Banerjee:2017hhz} 
 for the dark photon search,  $E_{A^\prime} \geq 0.5E_0$,  is associated 
 with the cut  for the millicharged particles search, 
 $E_{\chi} + E_{\bar{\chi}} \geq 0.5 E_0$. 
 In Fig.~\ref{mQReachFig} we show relevant excluded area
by blue shaded region.     

\section{Combined expected bounds from NA64e and NA64$\mu$
\label{MillBounds}}
In this section we estimate expected bounds on charge 
of millicharged particles using the results of the 
Section~\ref{BASIC} and Section~\ref{Elimits}. For thin target with $L_T \ll X_0^\mu$ 
the millicharge yield, which originates  from the 
 $\mu Z \rightarrow \mu Z \chi \chi$  process,  can be estimated as
 \begin{equation}
 N_{\chi \chi} = N_{MOT} \times \frac{\rho \times N_A}{A}  \times L_T \times \sigma_{\chi \chi},
 \label{Nchichi1}
 \end{equation}
where $A$ is the atomic weight, $N_A$ is Avogadro's number, $\rho$ denotes the 
target density and $\sigma_{\chi \chi} $ 
is the the 
millicharged pair production cross-section (\ref{eNtoeNcc2}). 
 In our estimates we assume that  the muon beam energy is 
about 100 GeV and the muon flux is about $N_{MOT}=5\cdot 10^{13}$. We 
consider  lead target with thickness of $L_T=40X_0=20$ cm.
We neglect muon energy losses  in the lead target. 
Indeed, this approximation is  reasonable, because the muon energy attenuation reported
in~\cite{Chen:2017awl} is rather small for the beam energy range,
$\left\langle dE_\mu/dz \right\rangle \approx 12.7 \cdot 10^{-3}$ GeV/cm. 
We assume that  the energies of initial and final muons are known. In NA64 facility with muon beam it is
 assumed to utilize 
two, upstream and downstream, magnetic spectrometers allowing for 
precise measurements of momenta for incident and recoiled muons, 
respectively~\cite{Gninenko:2014pea}.      
The  missing energy signature 
can  be used for the search for pair produced millicharged particles in the millicharged production reaction
$\mu N \rightarrow  \mu N  \chi\chi $ in a way analogous to the case of 
dark photon searching.       

\begin{figure}[tbh!]
\begin{center}
\includegraphics[width=0.5\textwidth]{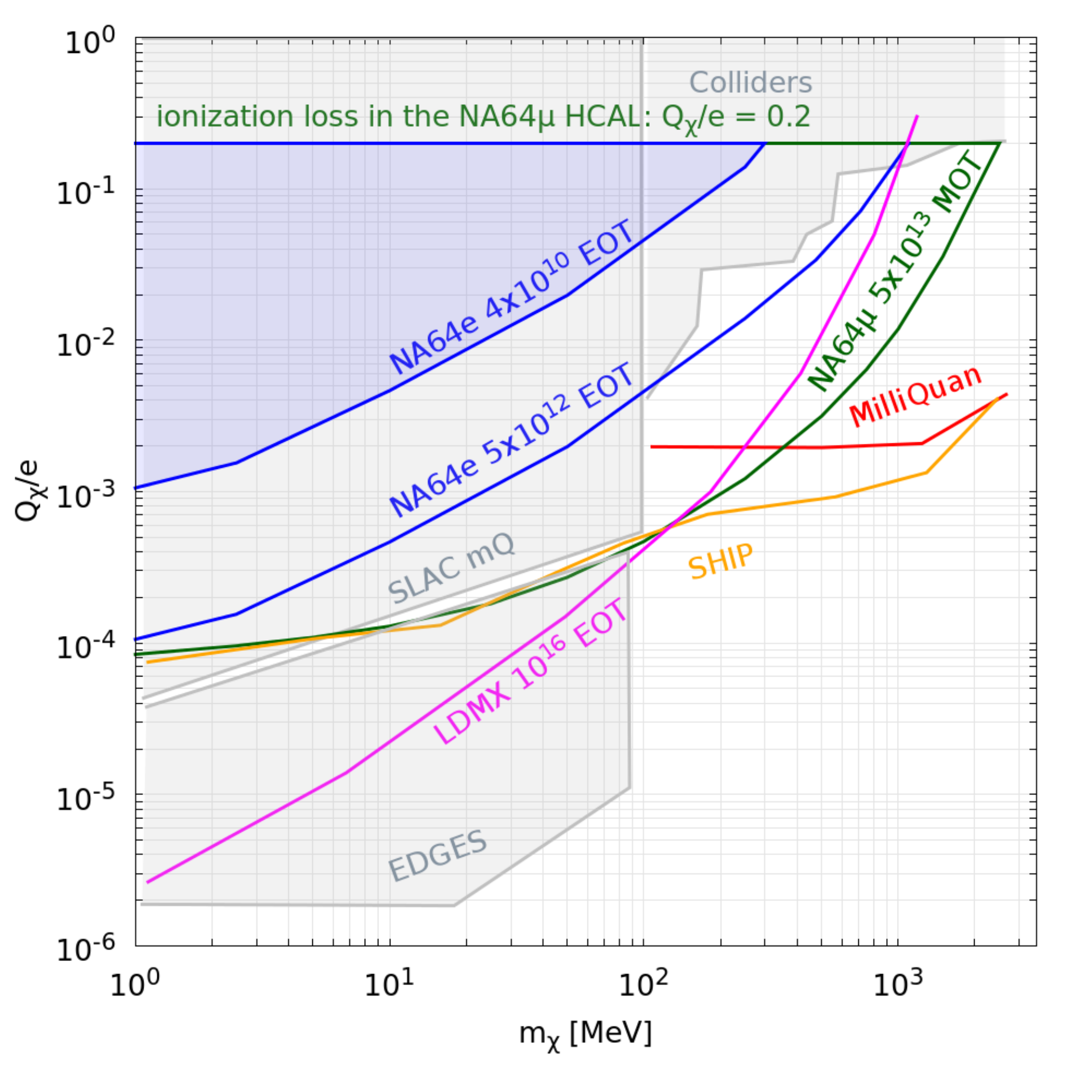}
\caption {Upper limits on the fractional electric charge $Q_\chi/e$ of the 
hypothetical millicharged fermions of mass $m_\chi$. The areas with the
 grey shading are the bounds excluded by SLAC~\cite{Prinz:1998ua}, 
 collider~\cite{Davidson:2000hf,Haas:2014dda}  and EDGES 
 experiment~\cite{Berlin:2018sjs,Barkana:2018qrx}. The projected limits are 
 shown by solid lines. In particular, the expected reaches for  SHIP and 
 MilliQuan   are taken from~\cite{Magill:2018tbb}. The sensitivity of LDMX 
  is based on {\tt MadGraph} missing momentum simulation with $16$ GeV  electron beam   on aluminium target~\cite{Berlin:2018bsc,Akesson:2018vlm}. 
  The blue shaded region is the bound experimentally excluded by NA64, 
  see e.g. Ref.~\cite{Banerjee:2017hhz}. The upper bound at 
  $Q_\chi/e = 0.2$ 
  corresponds to the (90\% CL) lower limit on the charge of $\chi$'s  
  above  which they are detected in the NA64 HCAL.}   
\label{mQReachFig}
\end{center}
\end{figure}

Using the formula (\ref{Nchichi1}) for the number of produced millicharged particles 
and the expression ~(\ref{eNtoeNcc2}) for the production cross section, we find 
the expected bound on millicharge $Q_{\chi}$.\footnote{We assume 
 background free regime, that looks reasonable, see~\cite{Gninenko:2014pea}.}
 We require 
$N_{\chi\chi} > 2.3$ that corresponds to  $90 \% CL$ exclusion limit
 on $Q_\chi/e$. In Fig.~\ref{mQReachFig} we show the
 expected reach of NA64 detector for 
 $N_{MOT}=5 \cdot 10^{13}$ muons and 
 $N_{EOT}= 5 \cdot 10^{12}$ electrons respectively, we assume that the  
 beam energy is $E_0=100$ GeV for both $e$- and $\mu$- modes. 
 
\begin{figure}[tbh!]
\begin{center}
\includegraphics[width=0.5\textwidth]{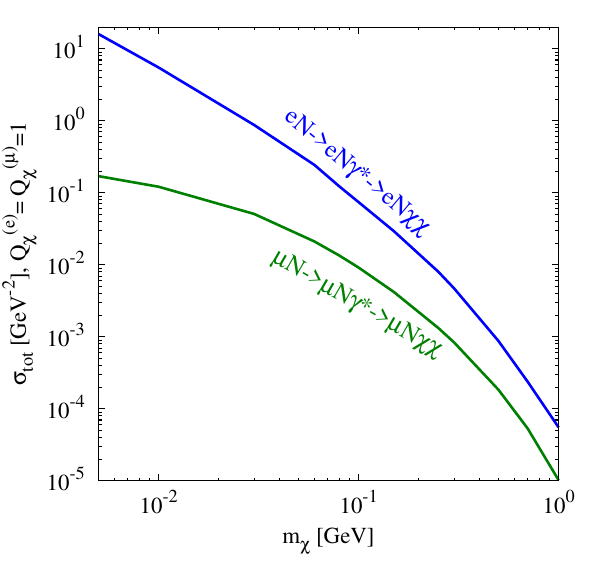}
\caption {The total production cross-sections of millicharged particles as a function their mass for 
the lead target and initial beam energy of $100$~GeV.
The green and blue curves show the cross-sections 
for muon and electron beam respectively. The cross-section are shown for $Q_\chi^{(\mu)} = Q_\chi^{(e)} =1$.}   
\label{CSeANDmuFig}
\end{center}
\end{figure}

It is instructive to compare qualitatively millicharge limits for muon and electron 
beam in order to understand  why  the expected bound from muon setup is enhanced at $m_\chi \gtrsim m_\mu $. 
Indeed, the ratio of the reaches can be naively 
approximated as follows
\begin{equation}
\frac{Q_\chi^{(e)}}{Q_\chi^{(\mu)}} \approx 
\left(\frac{L_{eff}^{(\mu)}}{L_{eff}^{(e)}} \cdot 
\frac{N_{MOT}}{N_{EOT}} 
\cdot \frac{\sigma^{(\mu)}_{\chi \chi}[Q_\chi^{(\mu)}=1]}{\sigma^{(e)}_{\chi \chi}[Q_\chi^{(e)}=1]} \right)^{1/2},
\label{QeOverQmu}
\end{equation}
where  labels $(\mu /e )$ specify a beam type and  
 $L_{eff}$ is the effective length of millicharge 
lepto-production in the lead target, namely 
$L_{eff}^{(\mu)} \approx  40 X_0$ and 
$L_{eff}^{(e)} \approx X_0 \approx 0.5~cm$. The latter means that for electron beam
the  millicharges are essentially produced in  the length 
$L \lesssim X_0$  of the target due to  the large electron energy loss. While 
the muons produce millicharges uniformly over the whole length of the 
target.
For beam energy $E_0 = 100$ GeV and millicharge masses 
$m_\chi \gtrsim 200$ MeV  the electron and muon
  cross-sections  scale respectively as \cite{Bjorken:2009mm}
\begin{equation}
\sigma^{(e)}_{\chi \chi} \sim \frac{Q_{\chi}^2}{(2 m_\chi)^2} 
\left(\ln \frac{1}{2} \left[ \frac{E_0}{  2 m_\chi}\right]^2 +\mathcal{O}(1) \right) , \qquad \sigma^{(\mu)}_{\chi \chi}\sim \frac{Q_{\chi}^2}
{ (2 m_\chi)^2} \left(  \ln 
\frac{1}{2}
\left[\frac{2 m_\chi}{m_\mu}\right]^2 +\mathcal{O}(1) \right),
\label{eANDmuCSlargeMass}
\end{equation}
where factor $1/2$ under the logarithms
 comes from the  integration of production
cross-section over the  missing energy range, $1/2 < E_{\gamma^*}/E_0<1$.
For $N_{MOT} = 5 \cdot 10^{13}$ and  
$N_{EOT} = 5 \cdot 10^{12}$ one 
has $Q_\chi^{(e)}/Q_\chi^{(\mu)} \approx 9$ at $m_\chi = 200$ MeV.
The corresponding result is shown in  Fig.~\ref{mQReachFig}. Indeed,
for $m_\chi = 200$ MeV we have $Q_\chi^{(e)}/e \approx 10^{-2}$ and 
$Q_\chi^{(\mu)}/e \approx 10^{-3}$. One can also estimate  
the ratio~(\ref{QeOverQmu}) for  relatively light millicharges $m_\chi \ll m_\mu$. 
In this case  muons produce the millicharges in the 
bremsstrahlung-like limit. The   cross-sections 
for both electron and muon beam can be  approximated as 
\begin{equation}
\sigma^{(e)}_{\chi \chi}\sim \frac{Q_{\chi}^2}{  (2 m_\chi)^2} \left(  \ln 
\frac{1}{2}
\left[\frac{2 m_\chi}{  m_e}\right]^2 +\mathcal{O}(1) \right),
\qquad 
\sigma^{(\mu)}_{\chi \chi}\sim \frac{Q_{\chi}^2}{  m_\mu^2} \left(  \ln 
\frac{1}{2}
\left[\frac{ m_\mu}{ 2 m_\chi}\right]^2 +\mathcal{O}(1) \right).
\label{CSemuCompSmallM}
\end{equation}
As a result   we get     $Q_\chi^{(e)} \gtrsim Q_\chi^{(\mu)}$ 
for $m_\chi \gtrsim 2$ 
MeV, see also  Fig.~\ref{mQReachFig}. In addition, 
to illustrate the increased sensitivity of NA64$\mu$,  in Fig.~\ref{CSeANDmuFig} we plot
 the total cross-sections~(\ref{eNtoeNcc2}) for muon and 
electron beam setup.
 One can see from 
Fig.~\ref{CSeANDmuFig} and Eq.~(\ref{CSemuCompSmallM}) that  for the small mass range $m_\chi < m_\mu$ 
the muon cross-section decreases slowly as $m_\chi$ approaches to $m_\mu$  in 
contrast to the electron beam case. On the other hand, for 
$m_\chi > m_\mu$ both the electron and the muon cross-sections scale as 
$1/m_\chi^2$ up to the logarithmic terms in~(\ref{eANDmuCSlargeMass}). 
Given that, one can gain the increased sensitivity~(\ref{QeOverQmu}) for 
the muon beam 
setup in the large mass range $m_\mu \gtrsim m_\mu$.
Namely,  at $m_\mu \simeq 200$~MeV one has $\sigma_{\chi\chi}^{(\mu)}[Q_\chi^{(\mu)}=1]/\sigma_{\chi\chi}^{(e)}[Q_\chi^{(e)}=1]\approx 1/5$,
 and, therefore we have
 $Q_\chi^{(e)}/Q_\chi^{(\mu)} \approx 9$ for $L_{eff}^{(\mu)}/L_{eff}^{(e)}\approx 40$ and $N_{MOT}/N_{EOT}\approx 10$, as outlined above.

The NA64e has an experience which indicates that the proposed search
for new physics in missing energy events with $3 \cdot 10^{11}$ EOT is essentially background
free due to the excellent detector capability for the identification of the initial and final state
in the high-energy electron interactions in the dump Ref.~\cite{NA64:2019imj}. The main potential sources of background are currently identified, and we do not expect a significant increase in background level after detector upgrade in 2021 run.

The design of NA64$\mu$ reported in Ref.~\cite{CERNSPSC}
is also based on the experience of NA64e. A preliminary study of the hadron contaminations and
detector hermeticity with the M2 muon beam shows that the total background level is conservatively
expected to be at the level $<10^{−12}$. Further improvements of this level are foreseen,  which, however,
will require modification of the beamline (e.g. adding of additional absorbers).

Before concluding, two remarks are in order. 
First, the NA64e experiment employed the optimized $100$~GeV electron
beam from the H4 beam line at the SPS. The beam has a maximal
intensity  $10^7$ e$^-$ per SPS spill of 4.8 s produced by the
primary $400$~GeV proton beam, and about $4000$ good spills per day.
Thus,  $\sim 120$ days are required  to accumulate $5\cdot10^{12}$ EOT  at H4 electron beamline.
Second, the intensity of $100$ GeV muons at the M2 beamline is a factor $10$ 
higher. Thus, about the same running time of 120 days is required to 
accumulate $5\cdot10^{13}$ MOT.
Here, the challenge might be to keep the overall efficiency still high at the intensity $\sim 10^8$ muons/spill, which however  seems possible~\cite{COMPASS}.

\section{Conclusions}

In this note we considered the prospects of the millcharged particle discovery 
at NA64 experiment at CERN in the MeV-GeV mass range of $\chi$. These 
weakly coupled to photons particles can be constrained 
by high energy collider experiments. 
 We studied  both cases of  
electron and muon 
beams at NA64 (NA64$e$ and NA64$\mu$).
We also calculate the relevant millicharge production cross-sections
in the equivalent photon approach. We analyse in detail 
 the corresponding parameter space 
 in the range of small, $m_\chi \ll m_\mu$, and large masses,  $m_\chi \gtrsim m_\mu$.
 We  found that NA64$\mu$ 
  can constrain the wider ranger of the millicharge 
 parameter space for $m_{\chi} \gtrsim m_{\mu}$  due to the increased accumulated statistic of muons and their small attenuation rate in the target. 
However, both NA64$e$ and NA64$\mu$ will be able to compete with current and forthcoming experimental facilities. Nevertheless,  
a more careful  Monte Carlo simulation of $\chi$ production at 
 NA64 is required to take into account the realistic response
 of the NA64 detector. We leave this task for future analysis.

\section{Acknowledgements}
We would like to thank P.~Schuster and G.~Lanfranchi for 
fruitful discussions.

\end{document}